\documentclass[submission,copyright,creativecommons]{eptcs}
\providecommand{\event}{MARS 2017} % Name of the event you are submitting to
\usepackage{underscore}           % Only needed if you use pdflatex.

\usepackage[framemethod=TikZ]{mdframed} %for boxes
\usepackage{amsmath} %for align in boxes % for matrices
\usepackage[export]{adjustbox} %for adjusting figures 1\textwidth and table
\usepackage{caption} %for subfigure
\usepackage{subcaption} %for subfigure
\usepackage{gensymb} %for degree symbol

\title{Towards Probabilistic Formal Modeling of \\ Robotic Cell Injection Systems}
\author{Muhammad Usama Sardar \qquad\qquad\qquad Osman Hasan
\institute{School of Electrical Engineering and Computer Science (SEECS)\\
National University of Sciences and Technology (NUST)\\
Islamabad, Pakistan}
\email{\{usama.sardar,osman.hasan\}@seecs.nust.edu.pk}
}
\def\titlerunning{Towards Probabilistic Formal Modeling of Robotic Cell Injection Systems}
\def\authorrunning{M.U. Sardar \& O. Hasan}
\begin{document}
\maketitle
\begin{abstract}
Cell injection is a technique in the domain of biological cell micro-manipulation for the delivery of small volumes of samples into the suspended or adherent cells. It has been widely applied in various areas, such as gene injection, in-vitro fertilization (IVF), intracytoplasmic sperm injection (ISCI) and drug development. However, the existing manual and semi-automated cell injection systems require lengthy training and suffer from high probability of contamination and low success rate. In the recently introduced fully automated cell injection systems, the injection force plays a vital role in the success of the process since even a tiny excessive force can destroy the membrane or tissue of the biological cell. Traditionally, the force control algorithms are analyzed using simulation, which is inherently non-exhaustive and incomplete in terms of detecting system failures. Moreover, the uncertainties in the system are generally ignored in the analysis. To overcome these limitations, we present a formal analysis methodology based on probabilistic model checking to analyze a robotic cell injection system utilizing the impedance force control algorithm. The proposed methodology, developed using the PRISM model checker, allowed to find a discrepancy in the algorithm, which was not found by any of the previous analysis using the traditional methods.
\end{abstract}

\section{Introduction}
\graphicspath{{images/}} %The path is relative to the current working directory.
Biological cell micro-manipulation primarily involves operating a cell by injection or removal of the samples \cite{Fleming2012}. 
In this domain, cell injection is a procedure where a small amount of material, such as protein, DNA, sperm or bio-molecules, is inserted into a specific location of suspended or adherent cells \cite{Kallio-JM2006}.
It has been widely adopted since early 1900s in gene injection \cite{Kuncova-EMBS2004}, in-vitro fertilization (IVF) \cite{Sun-IJRR2002}, intracytoplasmic sperm injection (ISCI) \cite{Yanagida-HR1999}, and drug development \cite{Nakayama-FS1998}.
For instance, it is used in ICSI to inject an immobilised sperm into the cytoplasmic of a mature egg to enable fertilisation. Similarly, it enables cellular biology researchers to inject drugs into a cell and observe the implications at the cellular level.

The cell injection process requires high precision movement of the micro-pipette to accurately perform the delicate tasks of puncture, penetration and deposition and thus the injection force is the key factor in the survivability of the injected cells \cite{Huang-TASE2009}. Accurate measurement and control of the injection force is of vital importance because even a minute excessive force can result in missing the desired deposition point \cite{FaroquePhD2016} and thus lead to the failure of the injection task \cite{Huang-CRB2006}. It may also damage the membrane or tissue of the cell \cite{Huang-CRA2007}. On the other hand, if the injection force is insufficient, it will not pierce the cell membrane \cite{FaroquePhD2016}. 

Conventionally, the analysis of the force control algorithm in cell injection systems is carried out using simulation-based techniques or by performing experiments. However, both of these methods inherently lack exhaustiveness \cite{fedeli2007} in terms of coverage of all the possible combinations of values of the parameters involved. This leads to another key challenge of such verification of cell injection systems, i.e., selection of test vectors. A random selection of test vectors cannot provide an assurance of correctness of the force control module of a  cell injection system since it might miss the interesting portion of the design space. Therefore, simulation-based and experiment-based verification of cell injection systems is incomplete with respect to error detection, i.e., all errors in a system cannot be detected \cite{hasan2015}. Moreover, there are a number of random factors affecting the results in real-time cell injection systems. For instance, internal disturbance such as plant uncertainty, external disturbance such as environmental effects and measurement noise of encoders can lead to unpredictable situations. However, they are generally ignored in the simulation-based verification. Thus, due to their inherent limitations and ignoring many random factors in cell injection systems, simulation-based verification cannot ascertain absolute correctness and may lead to failures of the injection task or destruction of the membrane or tissue of the cell.

To overcome the above-mentioned limitations, we propose to use probabilistic model checking \cite{baier2008} for a rigorous and accurate analysis of cell injection systems while incorporating the random factors. Probabilistic model checking is an advanced formal verification technique for modeling and analyzing systems that exhibit probabilistic behavior. The key benefit of using probabilistic model checking in our approach is that it caters for incorporating the real-world disturbances and noise in the models and thus can be utilized for the quantitative analysis of cell injection systems. To formalize the system, we have chosen the PRISM model checker \cite{PRISM}  because of its rich semantics to allow specification of a wide variety of probabilistic systems. Moreover, it has support for automated analysis of a wide range of quantitative properties of the models. PRISM has been widely used to formally model and rigorously analyze functional, performance and reliability properties of a plethora of practical systems, including randomized distributed algorithms \cite{Kwiatkowska2012}, aerodynamics \cite{Sardar2016SATS}, security protocols \cite{Basagiannis2011} and biological systems \cite{Lakin2012}. The novel contributions of the paper are summarized below:
\begin{itemize}
	\item We present a formal model of the force control module of a cell injection system \cite{Huang-CRB2006}. To depict a real-world scenario, our formalization incorporates various probabilistic elements, such as internal and external disturbances and measurement noise. These randomized elements have a major effect upon the insertion force and thus survivability of cells and hence their inclusion in the model can provide very useful insights about the effectiveness of a given cell injection system.
	\item The proposed formalization allowed to find a discrepancy in the controller, i.e., image-based torque controller equation in the X-Y plane, which was not found by any of the previous analysis using the traditional methods. It illustrates the effectiveness and utilization of the proposed formalization.
	\item We have made our PRISM model available as open-source \cite{codelink} for download to facilitate further developments and analysis of the cell injection systems. 
\end{itemize}	

The rest of the paper is organized as follows: Section \ref{Sec:Prelim} provides a brief introduction to the PRISM model checker to facilitate the understanding of the rest of the paper. Section \ref{Sec:Modeling} explains the proposed  DTMC formalization of components of the cell injection system in PRISM along with the assumptions used in our model. 
It also presents the discrepancy found in the controller. 
Finally, Section \ref{Sec:Conclusion} concludes this paper.

\section{PRISM Model Checker}
\label{Sec:Prelim}
PRISM \cite{PRISM} is a probabilistic model checker, i.e., a software tool for the formal modelling and analysis of systems that exhibit random or probabilistic behaviour. It incorporates state-of-the-art symbolic data structures and algorithms. PRISM supports several types of probabilistic models, such as
discrete-time Markov chains (DTMCs),
continuous-time Markov chains (CTMCs) \cite{Kulkarni1995},
Markov decision processes (MDPs) \cite{Puterman1994},
probabilistic automata (PAs) \cite{Segala1995},
probabilistic timed automata (PTAs) \cite{beauquier2003} as well as extensions of these models with rewards (or costs), referred to as (discrete- or continuous-time) Markov reward models and priced PTAs.

The models of the systems are described using the \emph{PRISM language}, which is a simple, state-based language based on Alur's Reactive Modules formalism \cite{Alur1999}. The PRISM language primarily consists of modules and variables. A model is composed of a parallel composition of a set of modules that can interact with each other. A module consists of local variables and guarded commands. The values of these variables at any given time represent the state of the modules and the guarded commands mimic the behavior of the modules. The global state of the whole model is determined by the local state of all modules. The syntax of a PRISM command is as follows: 
\begin{mdframed}[skipbelow=0pt,skipabove=2pt]
	\begin{align}
		\label{CL-command}
		\texttt{[action] guard -> prob\_1 : update\_1 + ... + prob\_n : update\_n;}
	\end{align}
\end{mdframed}
\noindent
where \texttt{action} is the optional synchronization label, \texttt{guard} is a predicate over all the variables in the model (including those belonging to other modules), \texttt{update} represents the new values of the variables in the module and \texttt{prob} represents a probability (or rate) assigned to the corresponding transition, which the module can make if the guard is true.

In the context of cell injection systems, PRISM allows validation of the functional and performance aspects of cell injection system at an early stage of the development cycle. We argue that this kind of validation can play a vital role in preventing the propagation of design errors at later stages, thereby achieving safer, cheaper, and faster development of cell injection systems \cite{Kim2013}. 

\section{Proposed Formal Model}
\label{Sec:Modeling}
Due to the involvement of various random factors in cell injection systems, it is necessary to include the probabilistic considerations in their analysis and safety verifications. Therefore, we propose to use probabilistic model checking \cite{baier2008} for the formalization of cell injection systems. Probabilistic model checking has been widely used to model and rigorously analyze performance and reliability of a plethora of practical systems, such as communication and security protocols, randomised distributed algorithms, game theory, biological systems and many others which exhibit random or probabilistic behaviour. The formal model, developed using probabilistic model checker, can then be used to determine the efficiency of the given cell injection system. 
The key advantages of using probabilistic model checking in our approach include the support for the quantitative analysis while 
catering for the real-world factors of random nature. Moreover, unlike simulation-based verification, it provides a complete and accurate analysis of the given system and supports automatic verification of a wide range of properties of probabilistic systems.

\subsection{Modeling Approach}
\begin{figure}[b]
	\centering
	\includegraphics[width=0.85\textwidth,center]{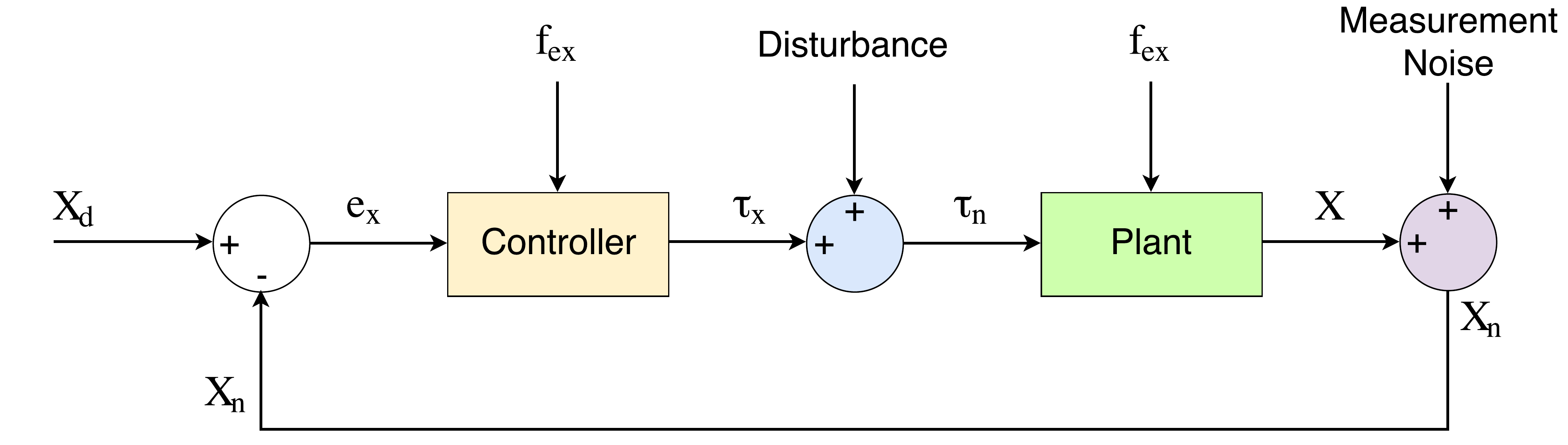}
	\caption{Closed-loop Model of the System}
	\label{Fig-System-Block}
\end{figure}

The force control module of cell injection systems is modeled as a closed-loop control system because of its better performance in terms of peak error compared to the open-loop system.
The model for the control of the pipette's trajectory in the X-coordinate is depicted in Fig. \ref{Fig-System-Block}. We use the term \emph{plant} for the robotic cell injection to distinguish it from the overall control system. The desired position of the pipette on the X-axis is represented by $X_d$ while $X_n$ denotes the position of the pipette on the X-axis, as given by the encoder. The difference $e_x=X_d-X_n$ represents the position/trajectory error in X-axis. The injection controller adjusts the torque $\tau _x$ input to the driving motor based on the error $e_x$ and the current value of the external force $f_{ex}$ applied to the actuators. The internal and external disturbances are modeled as additive noise. The resulting torque $\tau _n$ is provided to the plant, which computes the actual position of the pippete $X$. The measurement noise due to various factors, such as encoder noise, fabrication variation, calibration error is modeled as additive noise to $X$, resulting in $X_n$, which is fed back to the controller. 

\subsection{Proposed Formalization}
\begin{figure}[h]
	\centering
	\includegraphics[width=0.65\textwidth,center]{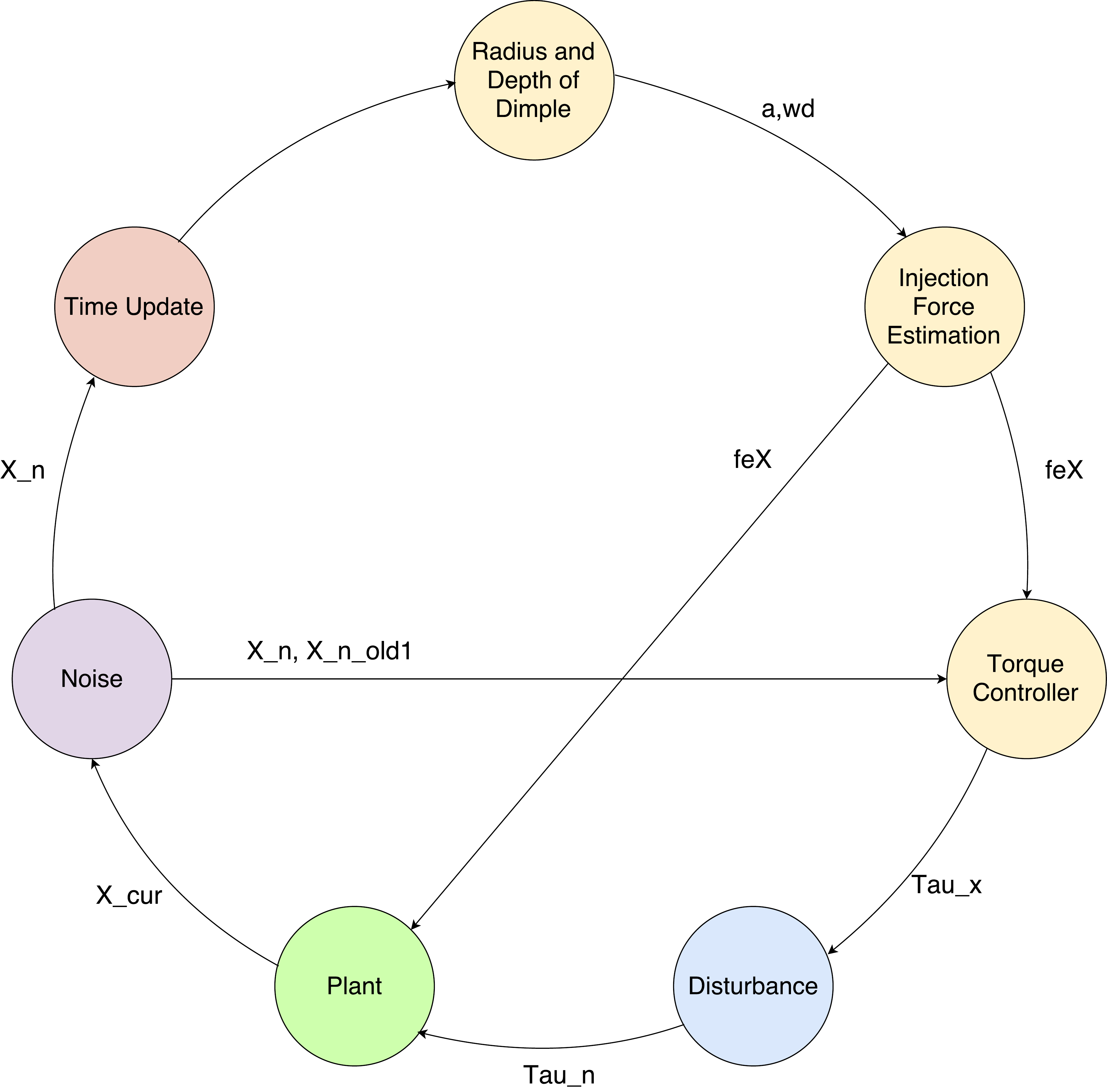}
	\caption{Finite Transition Model of the System}
	\label{Fig-FSM}
\end{figure}

We propose to utilize DTMC for the formalization because any physical system has to be discretized for practical implementation. In case of cell injection systems, the CCD camera that captures the images of the injected cell has a specific time interval. 
Moreover, there is a processing time to find the values of the radius \texttt{a} and depth \texttt{wd} of the dimple created. Based on these values, the value of force \texttt{feX} is estimated and is provided to the controller to compute the torque \texttt{Tau\_x}, as shown in Fig. \ref{Fig-FSM}. Torque \texttt{Tau\_n}, after addition of the internal and external disturbances, is applied to the motor in the plant. 
The actual position of the pipette \texttt{X\_cur} is corrupted by an additive noise to model the encoder measurement \texttt{X\_n}.
The resulting error \texttt{ex} is computed by the difference of desired and practical positions, and fed back to the controller. We assume that one time unit corresponds to the total time taken in this process. This process is repeated at each time instance during pre-piercing, piercing, injection and injector pulling out periods. After pulling out, the whole procedure is repeated for a new cell and thus the chain continues for a batch injection process. Hence, DTMC is the most appropriate underlying model for the analysis of these schemes and thus time is modeled as a discrete variable in our formalization.

\begin{figure}[]
	\centering
	\includegraphics[width=0.65\textwidth,center]{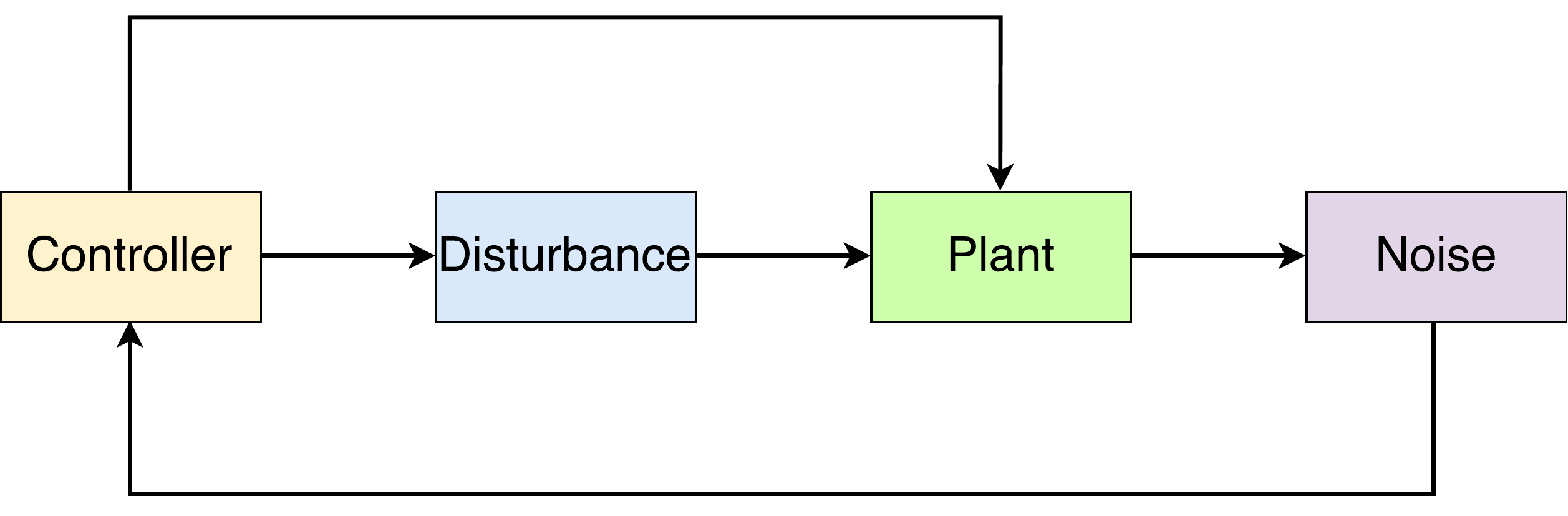}
	\caption{PRISM Modules and their Interaction}
	\label{Fig-PRISM-Modules}
\end{figure}

The PRISM model consists of 4 major modules, i.e., \texttt{plant, controller, disturbance} and \texttt{noise}, and their interaction is shown in Fig. \ref{Fig-PRISM-Modules}. The \texttt{plant} module models the dynamics of the plant. The \texttt{controller} module implements the image-based injection controller. 
The \texttt{disturbance} and \texttt{noise} modules present our formalization of the internal and external disturbance, as well as measurement noise. 
\subsection{Formal Model of the Plant}
\label{SubSec-Plant}
\begin{figure}[]
	\begin{subfigure}[t]{0.5\textwidth}
		\centering
		%\hspace{-2cm}
		\includegraphics[width=1\textwidth,center]{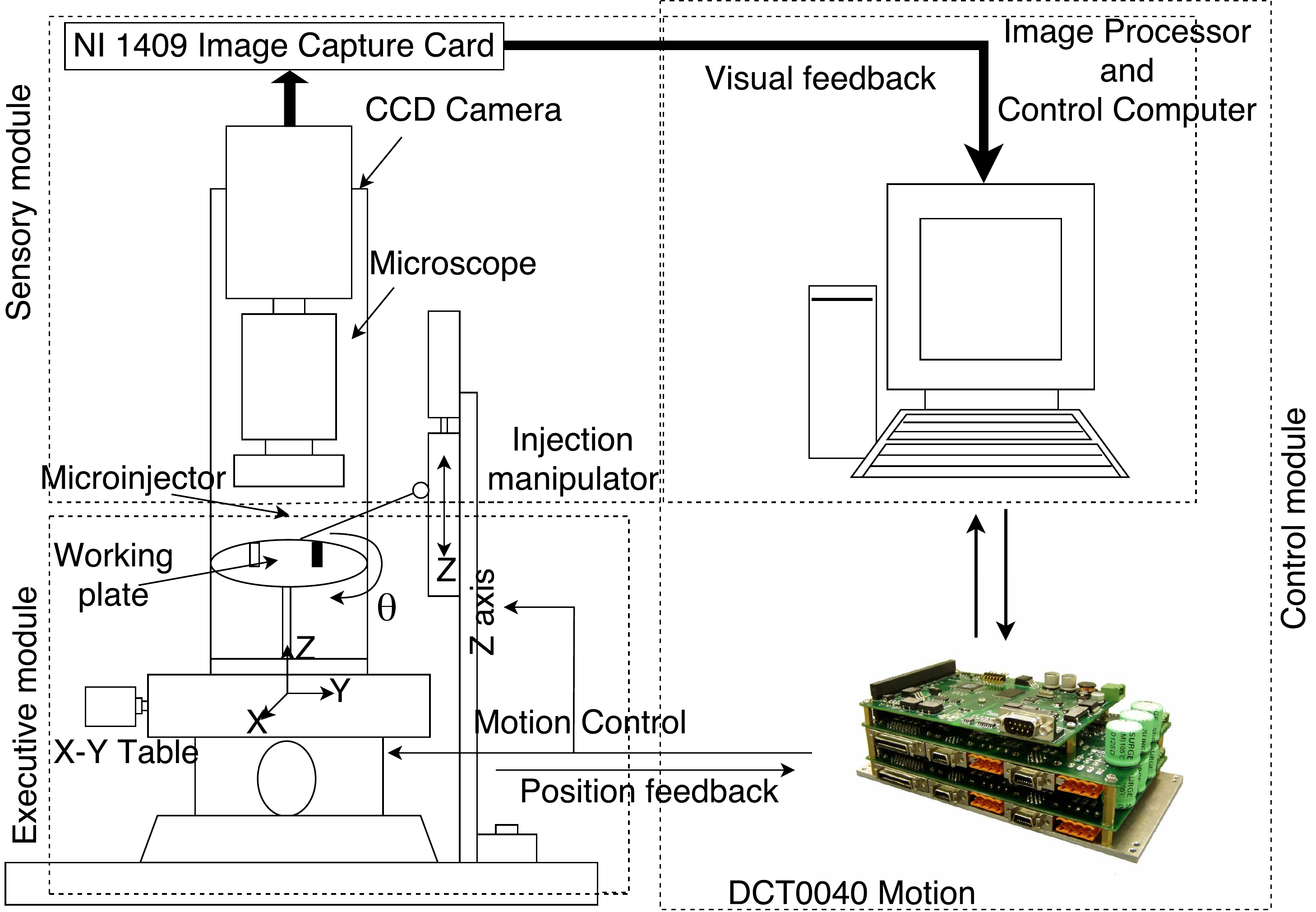}
		\caption{Modules of the Plant \cite{Huang-CRB2006}}
		\label{Fig-Modules}
	\end{subfigure}%	
	%~
	\begin{subfigure}[t]{0.5\textwidth}
		\centering
		\includegraphics[width=0.6\textwidth,center]{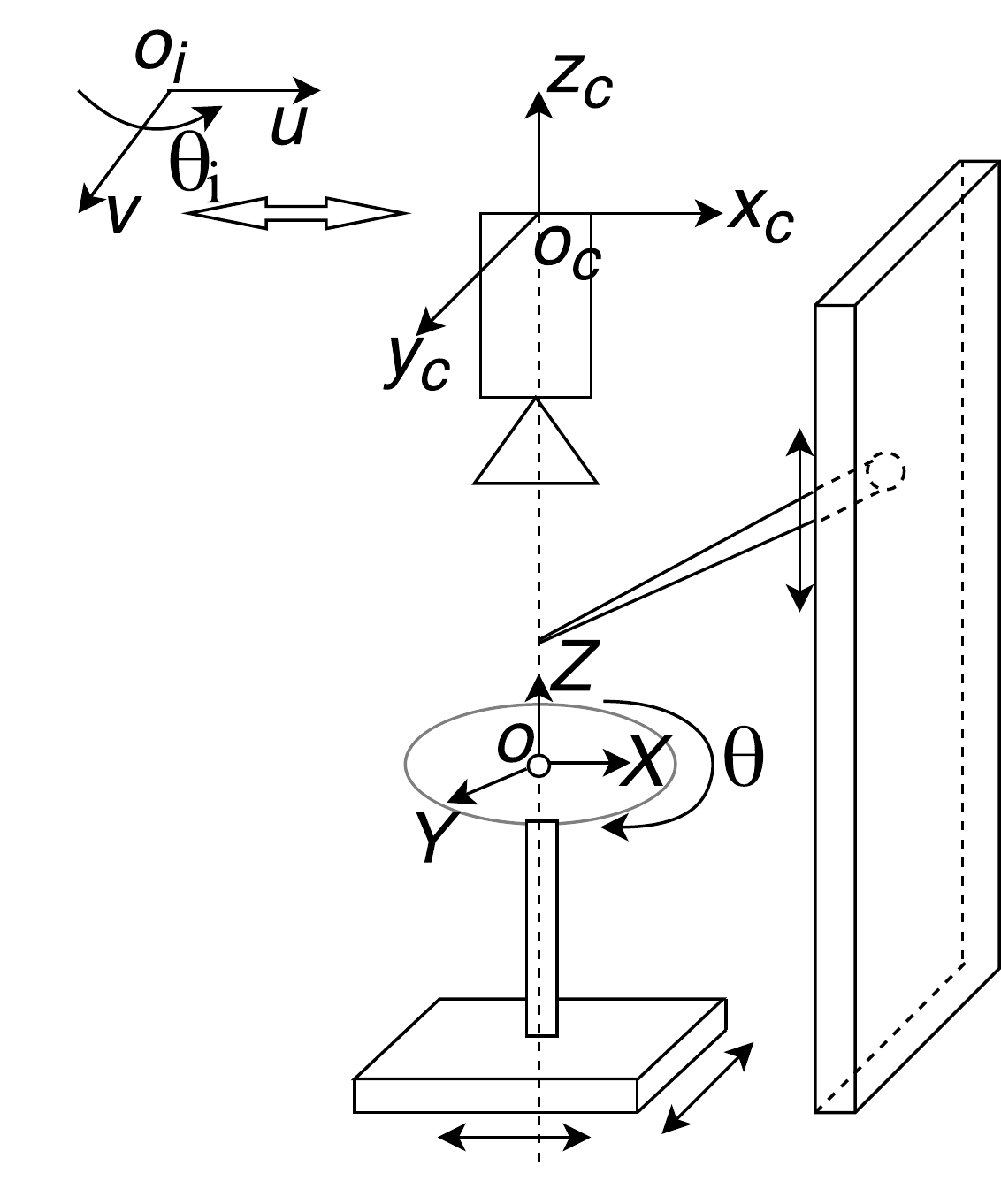}
		\caption{Configuration of the Plant \cite{Huang-CRB2006}}
		\label{Fig-Configuration}
	\end{subfigure}
	\caption{Overview of the Plant}
	\label{Fig-Overview}
\end{figure}

The major modules of the plant are illustrated in Fig. \ref{Fig-Modules}. We define $o-XYZ$ as the stage coordinate frame whose origin $o$ is located at the center of the working plate, and $o_i-uv\theta_i$ as the coordinate frame in the image plane, as shown in Fig. \ref{Fig-Configuration}. The relationship between image coordinates $[u,v]^T$ and stage coordinates $[X,Y]^T$ is given by \cite{Huang-CRB2006}:

\begin{equation}
\begin{bmatrix} 
u \\
v
\end{bmatrix}
=
\begin{bmatrix} 
f_xCos\alpha & f_xSin\alpha \\
-f_ySin\alpha & f_yCos\alpha 
\end{bmatrix}
\begin{bmatrix} 
X \\
Y
\end{bmatrix}
+ 
\begin{bmatrix} 
f_xd_x \\
f_yd_y 
\end{bmatrix}
= T 
\begin{bmatrix} 
X \\
Y
\end{bmatrix}
+ 
\begin{bmatrix} 
f_xd_x \\
f_yd_y 
\end{bmatrix}
\end{equation}

\noindent
where $f_x=\lambda/\delta_u$ and $f_y=\lambda/\delta_v$ denote the display resolutions of the vision system in the two directions, $\lambda$ is the magnification factor of the microscope objective, $\delta_u$ and $\delta_v$ are the actual distances between two adjoining pixels in the CCD sensor u-v frame, $d_x$ and $d_y$ represent the displacements between origins of the stage frame and the camera frame, and $\alpha$ is the angle between the stage frame and the camera frame. For simplicity, the transformation matrix between image frame and stage frame is denoted by $T \in  R^{2x2}$.

The Euler's dynamics model considers only the forces at the boundaries of the object under investigation while Lagrange's dynamics takes into account the details. So we use Lagrange's equation of the motion \cite{Thorby2008} and the dynamics in the X-Y plane is given as \cite{Huang-CRB2006}: 

\begin{equation}
\label{Eq-Lagrange}
\begin{bmatrix} 
m_x+m_y+m_p   & 0 \\
0             & m_y+m_p
\end{bmatrix}
\begin{bmatrix} 
\ddot{X} \\
\ddot{Y}
\end{bmatrix}
+ N_{xy}'
\begin{bmatrix} 
\dot{X} \\
\dot{Y}
\end{bmatrix}
=
\begin{bmatrix} 
\tau_x \\
\tau_y 
\end{bmatrix}
-
\begin{bmatrix} 
f_{ex} \\
f_{ey} 
\end{bmatrix}
= \tau - f_e
\end{equation}

\noindent
where $m_x$ and $m_y$ represent the masses of the X and Y positioning tables, $m_p$ denotes the mass of the working plate, %$I_p$ is the inertia of the rotational axis and the working plate, 
$N_{xy}' \in  R^{2x2}$ 
denotes a diagonal matrix of the positioning table that reflects damping and viscous friction effects, $\tau=[\tau_x,\tau_y]^T$ is the per-unit torque input to the driving motors and $f_e=[f_{ex},f_{ey}]^T$ is the external force applied to the actuators during cell injection. The following values are obtained from the real-world experimental setup \cite{Huang-CRB2006}: 

	\begin{equation*}
	 N_{xy} = 
	\begin{bmatrix} 
	\frac{1.146}{\dot{X}}+(2.465e-0.03)\dot{X}-0.0479   & 0 \\
	0             & \frac{0.172}{\dot{Y}}+(1.473e-0.03)\dot{Y}+0.04023
	\end{bmatrix}
	kg \cdot m^2/s,  
	\end{equation*}
	\begin{equation*}
	M_{xy} = 
	\begin{bmatrix} 
	0.022180   & 0 \\
	0             & 0.011386
	\end{bmatrix}
	kg \cdot m^2, \lambda = 60 , \alpha = 0 \degree 
	\end{equation*}

\noindent
where $N_{xy}=N_{xy}'T^{-1}$. For the PULNiX TM-6701AN progressive scan CCD camera used in the plant, $\delta_u$ and $\delta_v$ are both $9 \mu m$ \cite{Pulnix2017}. Substituting all these values in Eq. \ref{Eq-Lagrange} for X-coordinate, we get the following second-order differential equation after simplification:

	\begin{equation}
	\label{Eq-Diff}
	0.022180 \ddot{X} + (2.465e-0.03)\dot{X}^2-0.0479\dot{X} = 1.5 \times 10^{-7} (\tau_x - f_{ex}) - 1.146
	\end{equation}

We propose to use the finite difference method \cite{Leveque2007} for discretizing the differential equation to its corresponding difference equation. Since forward and central difference operators require a future sample for computation of the derivative, we use the backward difference operators. We use the \textit{first-order accurate approximations} since higher-order difference schemes lead to considerable overhead cost. Moreover, higher-order terms are inaccurate, due to the noise effects inherent in any practical situation \cite{Ogata2010}. Representing the value of $X$ at time instant $t_i$ as $X_i$, the first and second derivatives of X using the backward difference method are given as:

	\begin{equation}
	\dot{X_i} = \frac {X_i - X_{i-1}} {\Delta t}
	%= \frac{dX}{dt}|_i
	\end{equation}
	\begin{equation}
	\ddot{X_i} = \frac {X_i - 2X_{i-1}+X_{i-2}} {(\Delta t)^2}
	\end{equation}
where $\Delta t$ represents the step size for time,  $X_{i-1}$ and $X_{i-2}$ denote the value of $X$ at time instants $t_{i-1}$ and $t_{i-2}$, respectively. Substituting these values in Eq. \ref{Eq-Diff} and ignoring higher-order terms, the solution for $X_i$ by manual arithmetic manipulation is given as follows:

	\begin{equation}
	X_i = \frac {(\frac{2}{(\Delta t)^2}- \frac{0.04749}{\Delta t}) X_{i-1} - \frac{1}{(\Delta t)^2}X_{i-2}+1.5 \times 10^{-7} (\tau_x - f_{ex}) - 1.146} {\frac{0.022180}{(\Delta t)^2}-\frac{0.04749}{\Delta t}}	
	\end{equation}
For simplicity, we assume $\Delta t=1$. After arithmetic simplification and rounding off the values to 6 decimal places, the final resulting equation is given as: 

	\begin{equation}
	\label{Eq-finalX}
	X_i = -77.143817 X_{i-1} + 39.510075 X_{i-2} - 5.926511 \times 10^{-6} (\tau_x - f_{ex}) + 45.278546
	\end{equation}
We formalize it in PRISM using the following command: 
\begin{mdframed}[skipbelow=0pt,skipabove=2pt]
	\begin{align}
	\label{CL-Xposition}
	\texttt{[] guard -> (X\_cur' = } &
	%\texttt{floor (a*X\_old1) + floor (b*X\_old2) } \\ &
	\texttt{ceil (p1*X\_old1 +  p2*X\_old2 } \\ &
	\nonumber
	\texttt{- p3*(TauX-feX) + p4));}
	\end{align}
\end{mdframed}
\noindent
where \texttt{X\_cur} denotes the current position of the pipette in the X-axis, \texttt{X\_old1} and \texttt{X\_old2} represent the position of the pipette in the X-axis at time instants $t_{i-1}$ and $t_{i-2}$, respectively. Moreover, \texttt{p1,p2,p3} and \texttt{p4} are defined as constants in PRISM and their values are obtained from Eq. \ref{Eq-finalX}. \texttt{TauX} represents the per-unit torque input to the driving motors and \texttt{feX} is the external force applied to the actuators during cell injection. We use the \texttt{ceil} function to round the final result to the nearest integer because PRISM does not support rational numbers for state variables. The \texttt{guard} denotes the condition for sequencing using a global variable \texttt{count}.

\subsection{Formal Model of the Controller}
\begin{figure}[]
	\centering
	\includegraphics[width=0.35\textwidth,center]{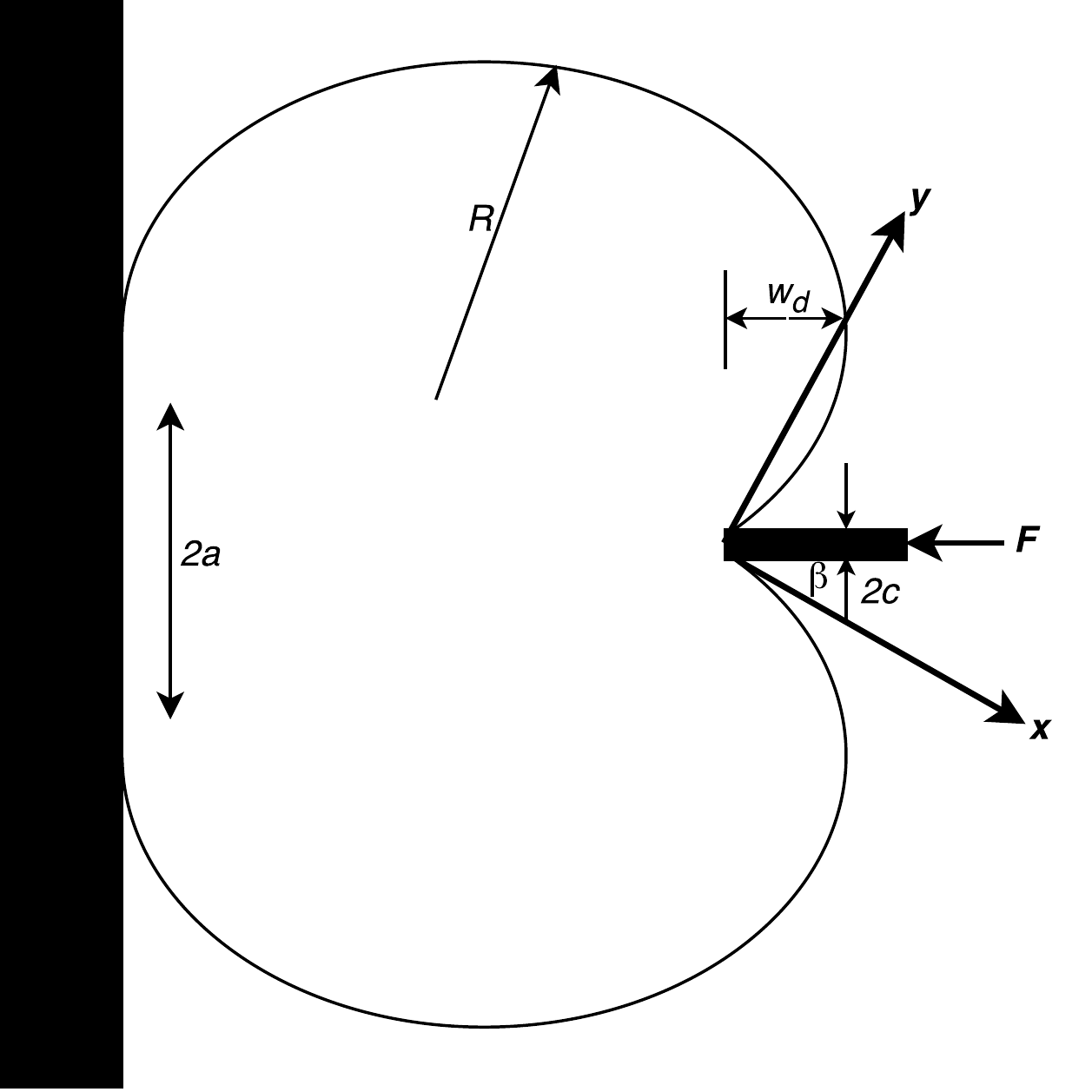}
	\caption{Bio-membrane Point Load Model \cite{Sun-TN2003}}
	\label{Fig-Biomembrane-model}
\end{figure}

We utilize the bio-membrane point load model \cite{Sun-TN2003}, as depicted in Fig. \ref{Fig-Biomembrane-model}, for the vision-based estimation of the injection force $F$, given as: 

	\begin{equation}
	\label{Eq-ForceF}
	F = \frac{2\pi Ehw_d^3}{a^2 (1-\gamma)} \frac{3-4\zeta ^2+\zeta ^4+2 ln \zeta^2}{(1-\zeta ^2)(1-\zeta ^2+ln\zeta ^2)^3}
	\end{equation}
where $E$ denotes the membrane elastic modulus, $h$ is bio-membrane thickness, $w_d$ is the depth of the dimple created due to injection of the pipette, $a$ is the radius of the dimple after injection of the pipette, $\gamma$ is the Poisson ratio, $\zeta=c/a$ where is $c$ is the radius of the pipette.
The formalization is presented below:
\begin{mdframed}[skipbelow=0pt,skipabove=2pt]
	\begin{align}
\texttt{[] guard -> } &
\nonumber
\texttt{(Force' = min(ceil((2*pi*EM*h*pow(wd,3))} \\ &
\nonumber
\texttt{*(3-4*pow(c/a,2)+pow(c/a,4)+2*log(pow(c/a,2),e))} \\ &
\nonumber
\texttt{/((pow(a,2)*(1-gamma))} \\ &
\nonumber
\texttt{*(1-pow(c/a,2))} \\ &
%\nonumber
\texttt{*pow((1-pow(c/a,2)+log(pow(c/a,2),e)),3))),Force\_max))} 
	\end{align}
\end{mdframed}	
where the variable \texttt{Force} represents the injection force, \texttt{EM} denotes the membrane elastic modulus, \texttt{h} is the  bio-membrane thickness, \texttt{wd} is the depth of the dimple created due to injection of the pipette, \texttt{a} is the radius of the dimple after injection of the pipette, \texttt{gamma} is the Poisson ratio and \texttt{c} is the radius of the pipette. \texttt{e} and \texttt{pi} are defined as constants.  The function \texttt{pow(i,j)} of PRISM computes $i$ to the power of $j$ and the function \texttt{log(i,j)} computes log of $i$ to the base $j$. The function \texttt{min(i,j)} returns the minimum of the two values $i$ and $j$, and is utilized to restrict the values within the maximum value of the force \texttt{Force\_max}.

To balance the force $F$, the external force $f_e$ applied to the actuator in the X-Y plane is given by: 
	\begin{equation}
	f_e = -F = 
	\begin{bmatrix}
	-F_x \\
	-F_y
	\end{bmatrix}
	= 
	\begin{bmatrix}
	-F \cdot cos \beta\\
	-F \cdot sin \beta
	\end{bmatrix}
	\end{equation}
where $\beta$ is the angle between injector and X-axis. For the system under consideration, $\beta=45$\degree \cite{Huang-CRB2006}. So, we have $f_e = 	\begin{bmatrix}
-\sqrt{2}/2 F \\
-\sqrt{2}/2 F
\end{bmatrix}$
N, where F is given by Eq. \ref{Eq-ForceF}.
This is formalized in PRISM using the following formula: 
\begin{mdframed}[skipbelow=0pt,skipabove=2pt]
	\begin{align}
	\label{CL-Force-feX}
	\texttt{formula feX = ceil (-Force / pow(2, 0.5));} 
	\end{align}
\end{mdframed}
where \texttt{feX} represents the x-component of the external force and \texttt{Force} is the injection force.

Using the contact space impedance force control \cite{Huang-CRB2006}, we have: 
	\begin{equation}
	\label{Eq-Contact}
	m \ddot{e} + b \dot{e} + ke = f_e
	\end{equation}
where $m$, $b$ and $k$ are mass, damping and stiffness constants, respectively, and $e=[X_d-X,Y_d-Y]^T$ is the position error. Solving for $[\ddot{X},\ddot{Y}]^T$ from Eq. \ref{Eq-Contact}, we get: 
	\begin{equation}
	\label{Eq-Xddot}
	\begin{bmatrix}
	\ddot{X} \\
	\ddot{Y}
	\end{bmatrix}
	= m^{-1} (m
	\begin{bmatrix}
	\ddot{X_d} \\
	\ddot{Y_d}
	\end{bmatrix}
	+ b \dot{e} + ke - f_e)
	\end{equation}
Now substituting Eq. \ref{Eq-Xddot} into Eq. \ref{Eq-Lagrange}, the image-based torque controller in the X-Y plane is given as: 
	\begin{equation}
	\label{Eq-Controller}
	\tau_{xy} = M_{xy} T 
	\begin{bmatrix}
	\ddot{X_d} \\
	\ddot{Y_d}
	\end{bmatrix}	
	+ M_{xy} T m^{-1} (b \dot{e} + ke - f_e)
	+ N_{xy} T
	\begin{bmatrix}
	\dot{X} \\
	\dot{Y}
	\end{bmatrix}	
	+ f_e
	\end{equation}	
while in the original paper \cite{Huang-CRB2006}, the controller is presented as: 
	\begin{equation}
	\label{Eq:Wrong-Eq}
	\tau_{xy} = M_{xy}  
	\begin{bmatrix}
	\ddot{X_d} \\
	\ddot{Y_d}
	\end{bmatrix}	
	+ M_{xy}  m^{-1} (b \dot{e} + ke - f_e)
	+ N_{xy} 
	\begin{bmatrix}
	\dot{X} \\
	\dot{Y}
	\end{bmatrix}	
	+ G_{xy} + f_e^d
	\end{equation}	
Although $G_{xy}=0$, so the term $G_{xy}$ does not make a difference in Eq. \ref{Eq:Wrong-Eq} but the transformations through matrix $T$ are missing and $f_e$ is wrongly presented as $f_e^d$. This discrepancy can lead to serious problems such as damaging membrane or tissue of the cell. Eq. \ref{Eq-Controller} is discretized using finite difference method, following a similar procedure as described in the plant (Section \ref{SubSec-Plant}), and the result is implemented in PRISM \cite{codelink}.

To mimic the image processing results, we model $a$ and $w_d$ differently in the four different time zones: pre-piercing, piercing, injection and pulling out. During the pre-piercing phase, the values of $a$ and $w_d$ increase by a non-decreasing factor due to the increasing velocity. During the piercing phase, the values of $a$ and $w_d$ decrease. While during the injection phase, they remain constant. Finally, during pulling out period, the factors decrease initially and then increase. Further details about our formalizations can be accessed from the source code \cite{codelink} of our model, which is shared under the Creative Commons Attribution-NonCommercial-ShareAlike 4.0 International License.

\subsection{Formal Model of Random Factors}
Various random factors play their role in affecting the results in a practical cell injection system and in many cases, the overall effect is significant and may result in serious problems. We classify the random factors as either disturbance or measurement noise. 
A disturbance may be generated within the system, called
\emph{internal} disturbance, or generated outside the system, known as \emph{external} disturbance \cite{Ogata2010}. Internal disturbances include plant uncertainties, such as electromagnetic effects of the components of the system, variability of operating point, wiring effect, variation in parameters of the process and distortion due to non-linear elements \cite{Dorf2011}. External disturbances are mainly caused by environmental effects, such as temperature and electromagnetic effects of components in the surrounding. For instance, the image degradation may occur at high temperatures \cite{Pulnix2017} and may result in error in the calculation of the parameters $a$ and $w_d$ of the biological cell. The measurement noise is due to the sensor error, including fabrication variation, lifetime of the sensor, and calibration error \cite{Dorf2011}. 

\begin{figure}[b]
	%\centering
	\begin{subfigure}[]{0.5\textwidth}
		\centering
		%\hspace{-2cm}
		\includegraphics[width=1\textwidth,center]{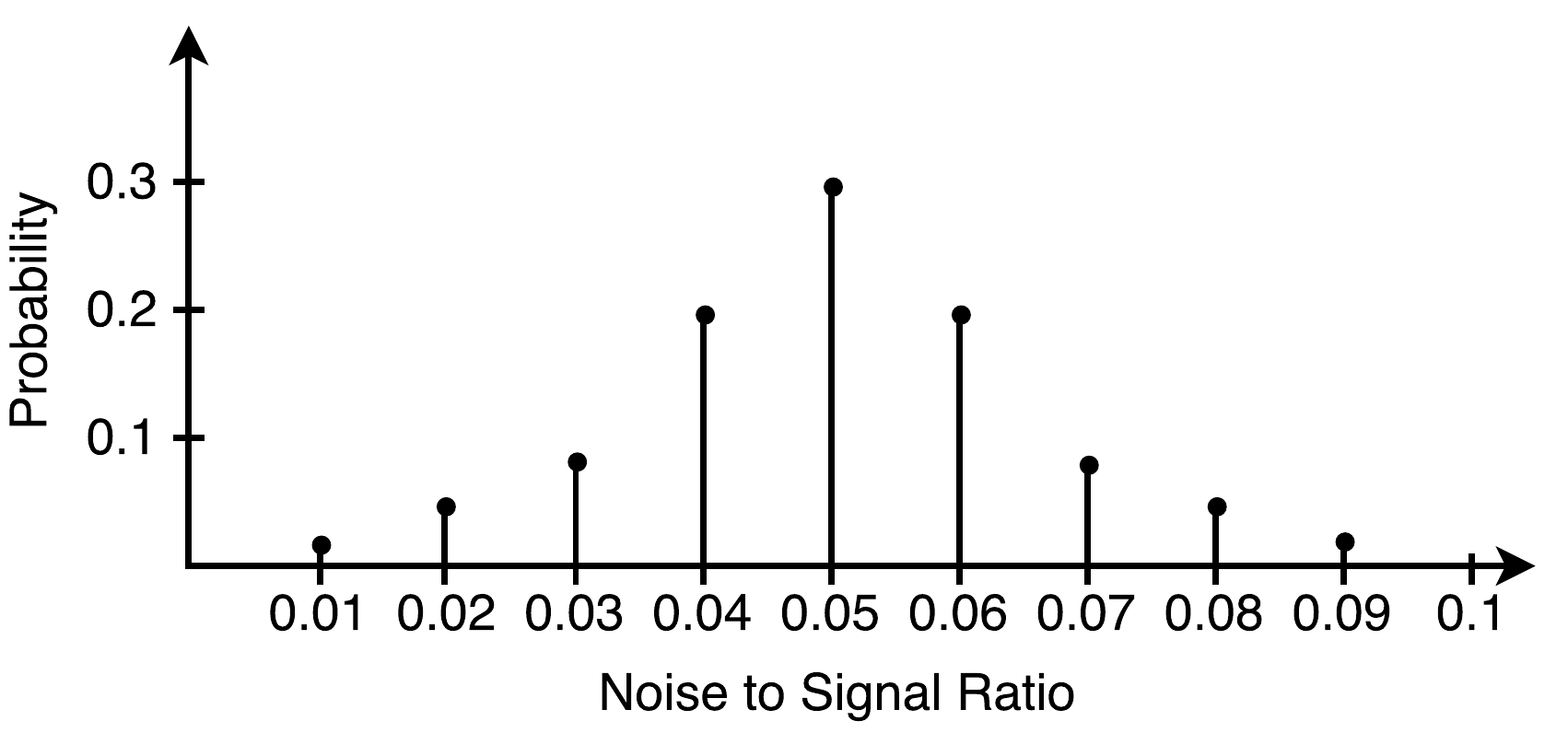}
		\caption{Model of the Disturbance}
		\label{Fig-Disturbance}
	\end{subfigure}%	
	%~
	\begin{subfigure}[]{0.5\textwidth}
		\centering
		\includegraphics[width=0.6\textwidth,center]{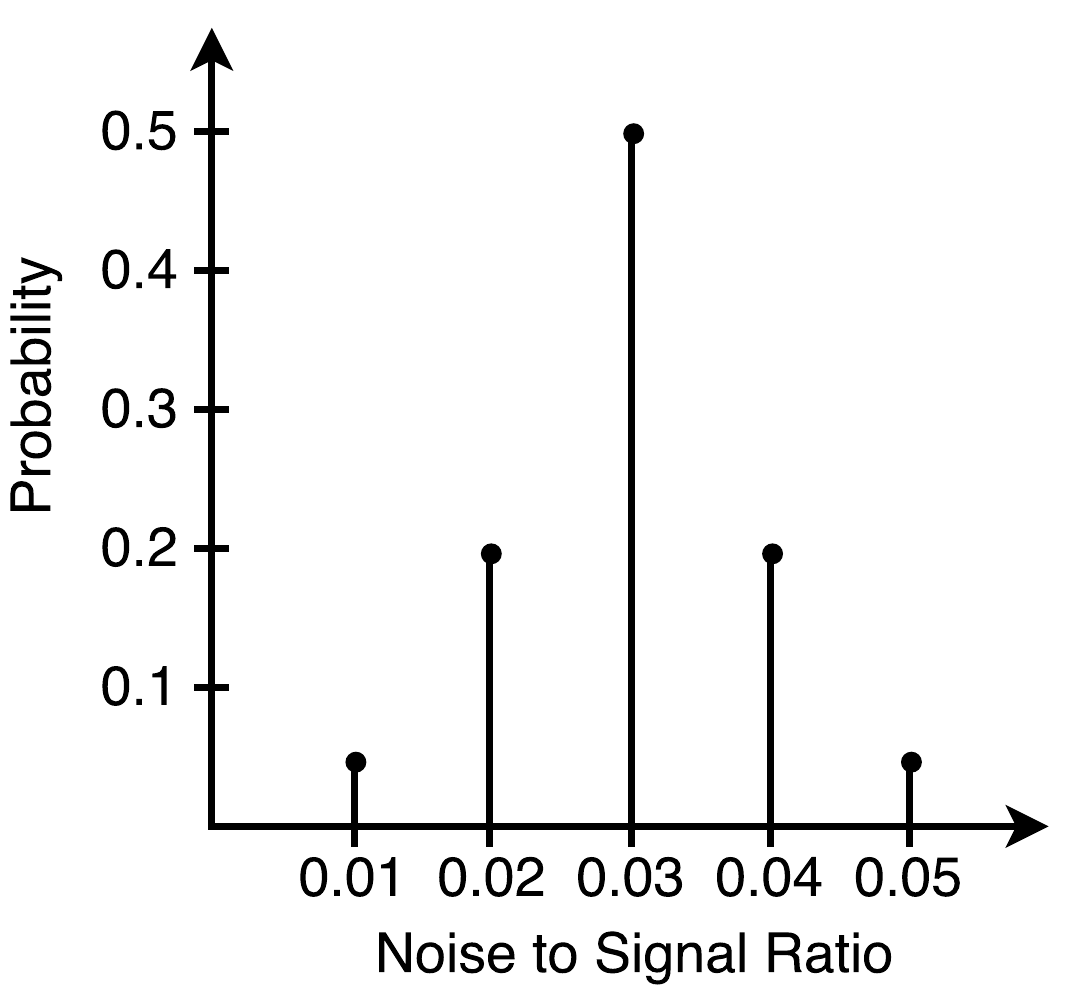}
		\caption{Model of the Measurement Noise}
		\label{Fig-Noise}
	\end{subfigure}
	\caption{Disturbance and Measurement Noise Models}
	\label{Fig-Noise-Models}
\end{figure}

We characterize the amount of noise by the noise-to-signal ratio \cite{Levine1999}. Our models of the disturbance and measurement noise are presented in Fig. \ref{Fig-Noise-Models}. Since the disturbance is generally greater than the measurement noise, we use the noise to signal ratios of 0.1 to 0.9 for disturbance, with the peak value at 0.05, as shown in Fig. \ref{Fig-Disturbance}. This is implemented in PRISM by the following command: 
\begin{mdframed}[skipbelow=0pt,skipabove=0pt]
	\begin{align}
	\label{CL-Distortion}
	\texttt{[] guard -> } &
	\nonumber
	\texttt{0.02:(Tau\_n' = ceil(Tau\_x + 0.01*Tau\_x))} +\\ &
	\nonumber
	\texttt{0.05:(Tau\_n' = ceil(Tau\_x + 0.02*Tau\_x))} +\\ &
	\nonumber
	\texttt{0.08:(Tau\_n' = ceil(Tau\_x + 0.03*Tau\_x))} +\\ &
	\nonumber
	\texttt{ 0.2:(Tau\_n' = ceil(Tau\_x + 0.04*Tau\_x))} +\\ &
	\nonumber
	\texttt{ 0.3:(Tau\_n' = ceil(Tau\_x + 0.05*Tau\_x))} +\\ &
	\nonumber
	\texttt{ 0.2:(Tau\_n' = ceil(Tau\_x + 0.06*Tau\_x))} +\\ &
	\nonumber
	\texttt{0.08:(Tau\_n' = ceil(Tau\_x + 0.07*Tau\_x))} +\\ &
	\nonumber
	\texttt{0.05:(Tau\_n' = ceil(Tau\_x + 0.08*Tau\_x))} +\\ &	
	\texttt{0.02:(Tau\_n' = ceil(Tau\_x + 0.09*Tau\_x));} 
	\end{align}
\end{mdframed}
\noindent 
where \texttt{Tau\_x} represents the per-unit torque from the controller while \texttt{Tau\_n} is the per-unit torque after addition of distortion. 

The noise to signal ratios of 0.1 till 0.5 are used for measurement noise, with peak value at 0.03, as shown in Fig. \ref{Fig-Noise}. This is implemented in PRISM as follows:
\begin{mdframed}[skipbelow=0pt,skipabove=2pt]
	\begin{align}
	\label{CL-Noise}
	\texttt{[] guard -> } &
	\nonumber
	\texttt{0.05:(X\_n' = ceil(X\_cur + 0.01*X\_cur))} +\\ &
	\nonumber
	\texttt{ 0.2:(X\_n' = ceil(X\_cur + 0.02*X\_cur))} +\\ &
	\nonumber
	\texttt{ 0.5:(X\_n' = ceil(X\_cur + 0.03*X\_cur))} +\\ &
	\nonumber
	\texttt{ 0.2:(X\_n' = ceil(X\_cur + 0.04*X\_cur))} +\\ &
	\texttt{0.05:(X\_n' = ceil(X\_cur + 0.05*X\_cur));}
	\end{align}
\end{mdframed}
\noindent 
where \texttt{X\_cur} denotes the position of the pipette on the X-axis and \texttt{X\_n} represents the position of the pipette on the X-axis after the addition of the measurement noise of encoder.

\newpage

\section{Conclusion}
\label{Sec:Conclusion}
Cell injection is the process of inserting a small volume of material, e.g., protein, DNA, sperm or biomolecules, into a specific location of suspended or adherent cells. It is widely used in drug development, toxicology, cellular biology research, transgenics, ICSI, and IVF. 
The injection force is the key factor in the success of the robotic cell injection process. We presented a formal modeling methodology based on the principles of probabilistic model checking for the injection force module of the cell injection systems in order to account for the disturbance and measurement noise.   
The presented formalization, using the PRISM model checker, helped in identifying a discrepancy in the force control algorithm of the robotic cell injection system. 
An important direction of future work is to carry out a detailed performance analysis of the given cell injection system on the basis of this formal model to verify properties, such as the probability of position and force errors exceeding certain thresholds. 

\section*{Acknowledgments}
We are thankful to  Khurram Ahmad for his help in elaborating the design of the controller.
This work was supported by a grant (number ICTRDF/TR\&D/2013/73) from National ICTR\&D Fund, Pakistan.

%\nocite{*} includes all references whether cited or not
\bibliographystyle{eptcs}
\bibliography{generic}
\end{document}